\documentclass[12pt,a4paper]{article}
\usepackage{epsfig}
\begin{document}
\title{Associated production of the top-pions and single top at hadron colliders}
\author{Chong-Xing Yue, Zheng-Jun Zong, Li-Li Xu, Jian-Xing Chen  \\
{\small Department of Physics, Liaoning  Normal University,
Dalian, 116029 P. R. China}
\thanks{E-mail:cxyue@lnnu.edu.cn}}
\date{\today}
\maketitle

\begin{abstract}
\hspace{5mm} In the context of topcolor assisted technicolor(TC2)
models, we study the production of the top-pions $\pi_{t}^{0,\pm}$
with single top quark via the processes $p\bar{p} \rightarrow
t\pi_{t}^{0}+X$ and $p\bar{p} \rightarrow t\pi_{t}^{\pm}+X$, and
discuss the possibility of detecting these new particles at
Tevatron and LHC. We find that it is very difficult to observe the
signals of these particles via these processes at Tevatron, while
the neutral and charged top-pions $\pi_{t}^{0}$ and
$\pi_{t}^{\pm}$ can be detecting via considering the same sign top
pair $tt\bar{c}$ event and the $tt\bar{b}$ (or $t\bar{t}b$) event
at LHC, respectively.

\vspace{1cm}


\end {abstract}

\newpage
\noindent{\bf I. Introducton}

The mechanism of electroweak symmetry breaking(EWSB) and origin of
the fermion mass remain unknown in elementary particle physics in
spite of the success of the standard model(SM) tested by high
energy experimental data. Hadron colliders, such as Tevatron and
Large Hadron Collider(LHC), are machines extremely well-suited to
study these problems. The LHC is expected to directly probe
possible new physics beyond the SM up to few $TeV$ and provide
some striking evidence of new physics, for instance of a light
Higgs boson, in its first months of operation[1].

Tevatron Run II has significant potential to discover a light SM
Higgs boson with mass up to about $M_{H}\leq 130GeV$[2]. The LHC
will have considerably capability to discover and measure almost
all the quantum properties of a SM Higgs boson of any mass[1].
However, if hadron colliders find evidence for a new scalar state,
it may not necessarily be the SM Higgs boson. Many alternative new
physics theories, such as supersymmetry, topcolor, and little
Higgs, predict the existence of new scalar or pseudo-scalar
particles. These new particles may have cross sections and
branching fractions that differ from those of the SM Higgs boson.
Thus, studying the production and decays of the new scalars at
hadron colliders will be of special interest.

Of particular interest to us is topcolor scenario[3], in which
there is an explicit dynamical mechanism for breaking electroweak
symmetry and generating the fermion masses including the heavy
quark mass. Thus, it is very attractive kind of models beyond the
SM. The presence of the physical top-pions $\pi_{t}^{0,\pm}$ in
low energy spectrum is an inevitable feature of the topcolor
scenario, regardless of the dynamics responsible for EWSB and
other quark mass. One of the most interesting features of
$\pi_{t}^{0,\pm}$ is that they have large Yukawa couplings to the
third-generation quarks and can induce the tree-level flavor
changing(FC) couplings[4].

In this paper, we will study the associated production of the
top-pions $\pi_{t}^{0,\pm}$ and single top quark via the
subprocesses $gc\rightarrow t\pi_{t}^{0}$ and $gb\rightarrow
t\pi_{t}^{\pm}$ and further discuss the possible signatures of
these new particles at the Tevatron and LHC experiments. Our
numerical results show that the top-pions can be significant
produced via these processes at LHC and their cross sections are
larger than those for the Higgs bosons $H^{0,\pm}$ predicted by
the minimal supersymmetric standard model(MSSM). These processes
can be used to probe the top-pions and distinguish the Higgs
bosons predicted by the MSSM from the top-pions predicted by the
topcolor scenario.

To completely avoid the problems arising from the elementary Higgs
field in the SM, various kinds of dynamical EWSB models have been
proposed, among which the topcolor scenario is attractive because
it can explain the large top quark mass and provide a possible
EWSB mechanism[3]. The topcolor-assisted technicolor(TC2)
models[5] are one kind of the phenomenologically viable models,
which has all essential features of the topcolor scenario. So, in
the rest of this paper, we will give our results in detail in the
context of the TC2 models.

This paper is organized as follows. Section II contains a short
summary of the relevant couplings to ordinary particles of the
top-pions $\pi_{t}^{0,\pm}$ in TC2 models. The anomalous top quark
coupling $tqv$ has contributions to the process $gq\rightarrow
t\pi_{t}^{0}$. Thus, the anomalous top quark coupling $tqv$ from
the FC interactions in TC2 models is also discussed in this
section. Sections III and IV are devoted to the computation of
production cross sections for the process $p\bar{p} \rightarrow
t\pi_{t}^{0}+X$ and $p\bar{p} \rightarrow t\pi_{t}^{-}+X$,
respectively. Some phenomenological analysis are also included in
these sections. Our conclusions are given in Sec.V.

\noindent{\bf II. The relevant couplings }

For TC2 models, the underlying interactions, topcolor
interactions, are nonuniversal and therefore do not possess the
Glashow-Iliopoulos-Maiani(GIM) mechanism. The nonuniversal gauge
interactions result in the new FC coupling vertices when one
writes the interactions in the mass eigen basis. Thus, the
top-poins $\pi_{t}^{0,\pm}$ can induce the new FC coupling
vertices. The couplings of $\pi_{t}^{0,\pm}$ to ordinary fermions,
which are related to our calculation, can be written as[4,5,6]:
\begin{eqnarray}
&&\frac{m_{t}}{\sqrt{2}F_{t}}\frac{\sqrt{\nu_{W}^{2}-F_{t}^{2}}}{\nu_{W}}
     [iK_{UR}^{tt}K_{UL}^{tt^{*}}\bar{t}\gamma^{5}t\pi_{t}^{0}+
     \sqrt{2}K_{UR}^{tt^{*}}K_{DL}^{bb}\bar{b}_{L}t_{R}\pi_{t}^{-}
+i\frac{m_{b}-m_{b}'}{m_{t}}\overline{b}\gamma^{5}b\pi_{t}^{0}\nonumber\\
&&+iK_{UL}^{tc^{*}}K_{UR}^{tt}\bar{c_{L}}t_{R}\pi_{t}^{0}
+\sqrt{2}K_{UR}^{tc^{*}}K_{DL}^{bb}\bar{b}_{L}c_{R}\pi_{t}^{-}+h.c.]
+\frac{m_{l}}{\sqrt{2}\nu_{W}}\bar{l}
     \gamma^{5}l\pi_{t}^{0},
\end{eqnarray}
where $\nu_{W}=\nu/\sqrt{2}=174GeV$, $l $ represents the lepton
$\tau$ or $\mu$, and $m_{b}'\approx0.1\varepsilon m_{t}$ is the
part of the bottom-quark mass generated by extended
technicolor(ETC) interactions. $K_{UL(R)}$ and $K_{DL(R)}$ are
rotation matrices that diagonalize the up-quark and down-quark
mass matrix $M_{U}$ and $M_{D}$ for which the
Cabibbo-Kobayashi-Maskawa(CKM) matrix is defined as $V
=K_{UL}^{+}K_{DL}$. To yield a realistic form of the CKM matrix V,
it has been shown that their values can be taken as[4]:
\begin{equation}
K_{UL}^{tt}\approx K_{DL}^{bb}\approx 1,\ \ \  K_{UR}^{tt}=
1-\varepsilon,\ \ \ K_{UR}^{tc}\leq
\sqrt{2\varepsilon-\varepsilon^{2}}\ .
\end{equation}
In the following calculation, we will take
$K_{UR}^{tc}=\sqrt{2\varepsilon-\varepsilon^{2}}$ and take
$\varepsilon$ as a free parameter, which is assumed to be in the
range of $0.01 \sim 0.1$[3,5].

\begin{figure}[htb]
\vspace{-7.4cm}
\begin{center}
\epsfig{file=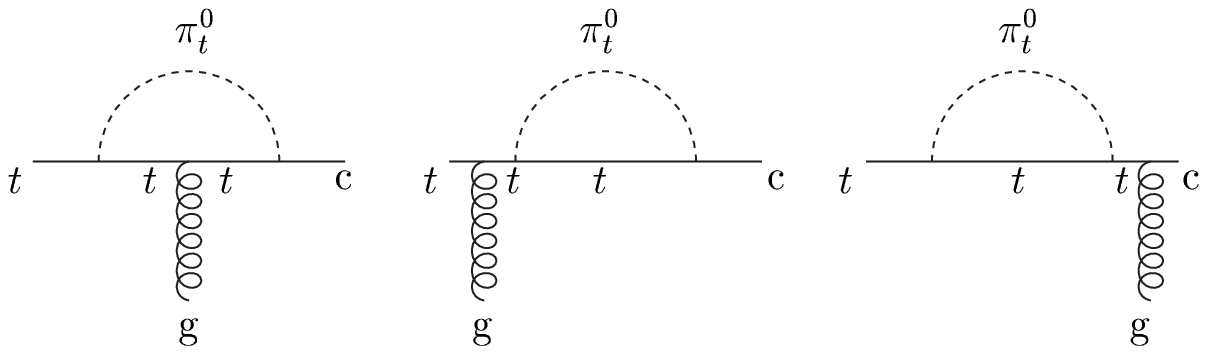,width=500pt,height=690pt} \vspace{-14.5cm}
\caption{Feynman diagrams for the contributions of $\pi_{t}^{0}$
to the anomalous top quark \hspace*{1.8cm}coupling $tcg$ }
\end{center}
\end{figure}

In the context of the SM, the anomalous top quark couplings $tqv$
(q=u- or c-quark and $v=Z $, $\gamma$ or $g$ gauge bosons), which
are arised from the FC interactions, vanish at the tree-level but
can be generated at one-loop level. However, they are strong
suppressed by the GIM mechanism, which can not produce observable
effects in the present and near future high energy experiments[7].
From Eqs.(1)and(2), we can see that the neutral top-pion
$\pi_{t}^{0}$ might generate the large top quark coupling $tcg$.
The relevant Feynman diagrams are shown in Fig.1. Similar to
Ref.[8], we can give the effective form of the anomalous coupling
vertex $tcg$:
\begin{equation}
\Lambda^{\mu}_{tcg}=ig_{s}\frac{\lambda^{a}}{2}[\gamma^{\mu}F_{1g}+p_{t}^{\mu}F_{2g}
+p_{c}^{\mu}F_{3g}].
\end{equation}
with
\begin{equation}
F_{1g}=\frac{1}{16\pi^{2}}[\frac{m_{t}}{\sqrt{2}F_{t}}\frac{\sqrt{\nu_{W}^{2}-F_{t}^{2}}}{\nu_{W}}]^{2}K_{UR}^{tc}
K_{UL}^{tt*}
(B_{0}+m_{\pi_{t}}^{2}C_{0}-2C_{24}+m_{t}^{2}(C_{11}-C_{12})-B_{0}^{*}-B_{1}^{'}),
\end{equation}
\begin{equation}
F_{2g}=2m_{t}\frac{1}{16\pi^{2}}[\frac{m_{t}}{\sqrt{2}F_{t}}\frac{\sqrt{\nu_{W}^{2}-F_{t}^{2}}}{\nu_{W}}]^{2}K_{UR}^{tc}
K_{UL}^{tt*}(C_{21}+C_{22}-C_{23}),
\end{equation}
\begin{equation}
F_{3g}=2m_{t}\frac{1}{16\pi^{2}}[\frac{m_{t}}{\sqrt{2}F_{t}}\frac{\sqrt{\nu_{W}^{2}-F_{t}^{2}}}{\nu_{W}}]^{2}K_{UR}^{tc}
K_{UL}^{tt*}(C_{22}-C_{23}+C_{12}),
\end{equation}
where $\lambda^{a}$ is the Gell-Mann matrix. The expressions of
the two and three-point scalar integrals $B_{n}$ and $C_{ij}$ are
[9]:
\begin{equation}
B_{0}=B_{0}(p_{g},m_{t},m_{t}),
\hspace{1cm}B_{0}^{*}=B_{0}(-p_{c},m_{\pi_{t}},m_{t}),
\end{equation}
\begin{equation}
B_{1}^{'}=B_{1}(-p_{t},m_{\pi_{t}},m_{t}),
\end{equation}
\begin{equation}
C_{24}=C_{24}(-p_{t},p_{g}, m_{\pi_{t}}, m_{t}, m_{t}),
\end{equation}
\begin{equation}
C_{ij}=C_{ij}(-p_{t}, -\sqrt{\hat{s}}, m_{\pi_{t}},m_{t},m_{t}),
\hspace{1cm}i, j=1, 2, 3.
\end{equation}

Certainly, the neutral top-pion $\pi_{t}^{0}$ can also generate
the anomalous top quark coupling $tug$ via the FC coupling
$\pi_{t}^{0}tu$. However, it has been argued that the maximum
flavor mixing occurs  between the third generation and the second
generation, and the FC coupling $\pi_{t}^{0}\bar{t}u$ is very
small which can be neglected[4]. Hence we will ignore the
contributions of the $tug$ coupling to the process
$p\bar{p}\rightarrow t\pi_{t}^{0}+X$ in the following discussions.

Similar to the neutral top-pion $\pi_{t}^{0}$, the charged
top-pions $\pi_{t}^{\pm}$ can generate the anomalous top quark
coupling $tcg$ via the FC couplings $\pi_{t}^{\pm}bc$. However,
compared with those of $\pi_{t}^{0}$, the contributions of
$\pi_{t}^{\pm}$ to the $tcg$ coupling are approximately suppressed
by the factor $m_{b}/m_{t}$, which can be safely neglected.

\noindent{\bf III. Associated production of the neutral top-pion
$\pi_{t}^{0}$ and single top quark}

\begin{figure}[htb]
\vspace{-4.4cm}
\begin{center}
\epsfig{file=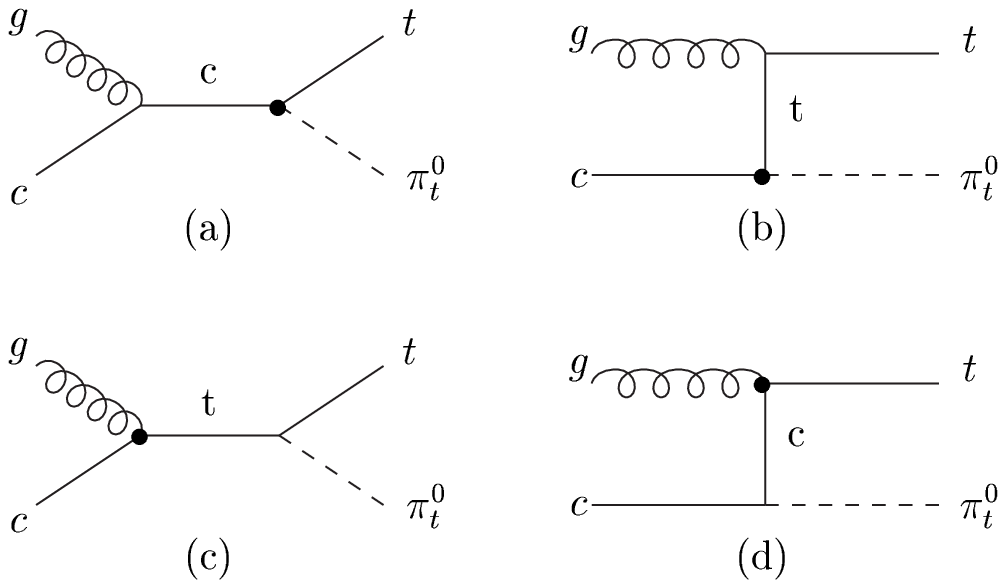,width=550pt,height=800pt} \vspace{-18.5cm}
\hspace{1cm} \vspace{-0.6cm} \caption{Feynman diagrams for the
process $g c \rightarrow t \pi_{t}^{0}$ } \label{ee}
\end{center}
\end{figure}

From above discussions, we can see that, due to the existence of
the FC couplings, the neutral top-pion $\pi_{t}^{0}$ can be
generated via the subprocess $gc\rightarrow t\pi_{t}^{0}$ at
hadron colliders, as shown in Fig.2. Fig.2(a) and Fig.2(b) come
from the FC coupling $\pi_{t}^{0}\bar{t}c$, while Fig.2(c) and
Fig.2(d) come from the anomalous top quark coupling $tcg$.
Although the strength of the coupling $tcg$ is very smaller than
that of the coupling $gc\bar{c}$ or $gt\bar{t}$, we can not ignore
the contributions of Fig.2(c) to the subprocess $gc\rightarrow
t\pi_{t}^{0}$ being large $\pi_{t}t\bar{t}$ coupling. For
Fig.2(d), it is not this case. The $\pi_{t}^{0}c\bar{c}$ coupling
is very small and thus the contributions of Fig.2(d) to the
subprocess $gc\rightarrow t\pi_{t}^{0}$ can be neglected. We have
confirmed this expectation through explicit calculation.

To obtain numerical results, we need to specify the relevant SM
input parameters. These parameters are $m_{t}=178GeV$ and
$\alpha_{s}(m_{t})=0.118$[10]. Through out this paper, we neglect
the charm quark mass and use CTEQ6L parton distribution functions
with scale $\mu=2m_{t}$[11]. The limits on the top-pion mass
$m_{\pi_{t}}$ can be obtained via studying its effects on various
observables[3]. It has been shown that $m_{\pi_{t}}$ is allowed to
be in the range of a few hundred GeV depending on the models. As
numerical estimation, we will assume that the value of the
top-pion mass $m_{\pi_{t}}$ is in the range of 200GeV $\sim$
500GeV.

\begin{figure}[htb]
\vspace{-0.4cm} \centering {
\label{fig:subfig:a}
\includegraphics[width=3.1in]{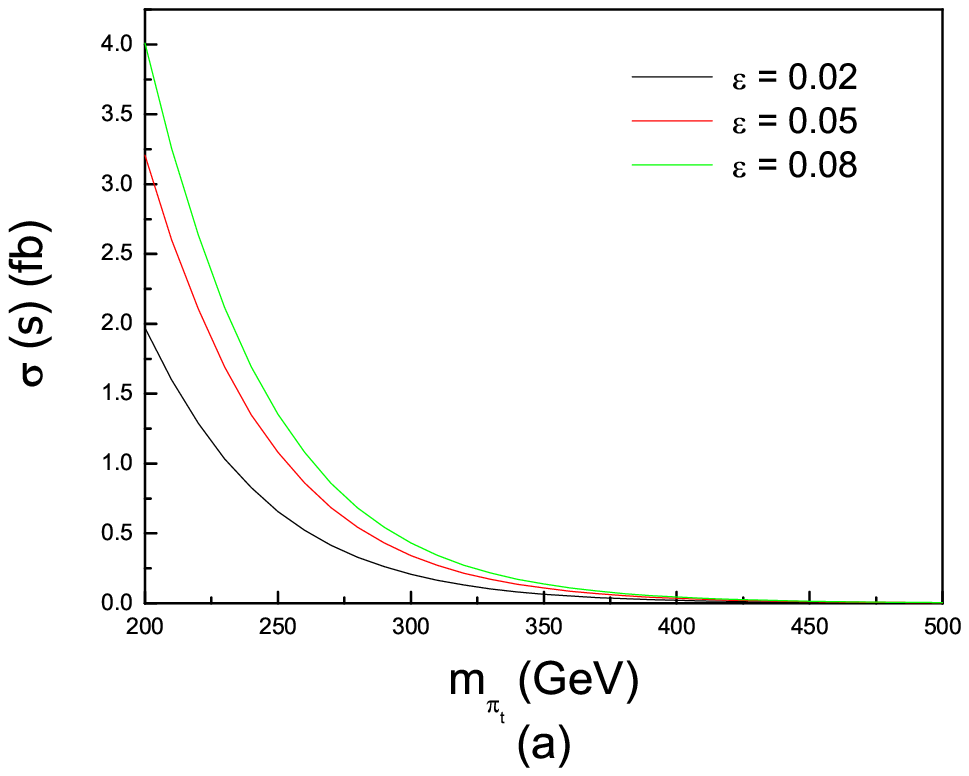}}
\hspace{-0.5cm} {
\label{fig:subfig:b}
\includegraphics[width=3.1in]{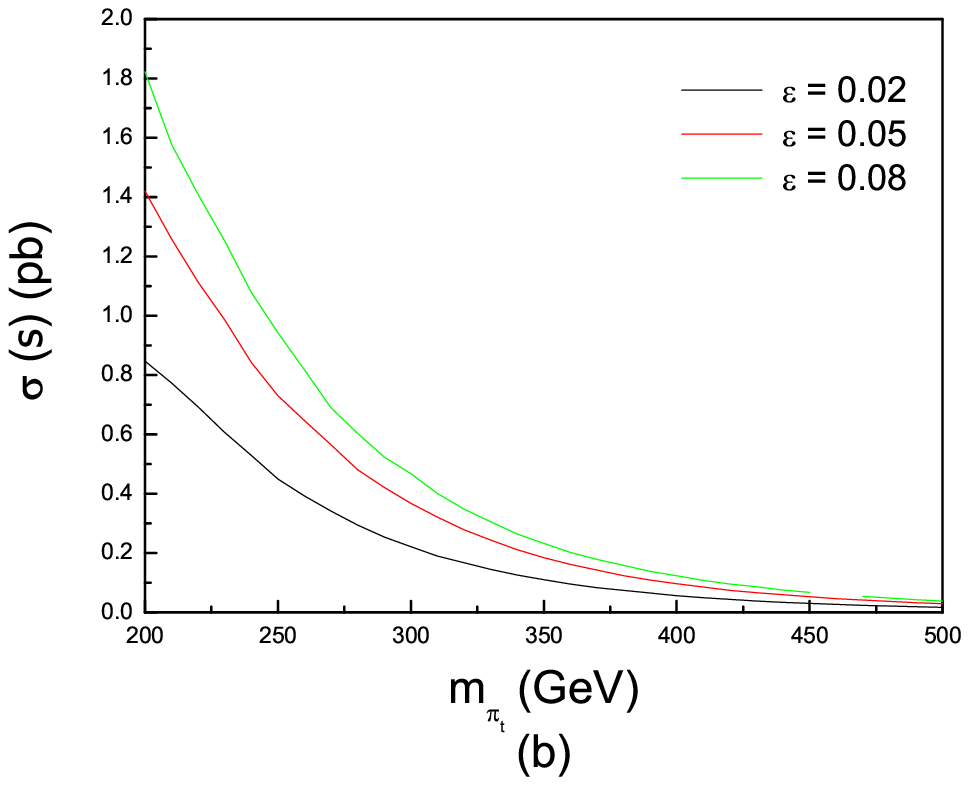}}
\caption{\footnotesize The cross section $\sigma(s)$ of
$t\pi_{t}^{0}$ production as function of $m_{\pi_{t}}$ for three
values of the parameter \hspace*{1.8cm} $\varepsilon$ at the
Tevatron with $\sqrt{s}=1.96TeV$(a) and the LHC with
$\sqrt{s}=14TeV$(b).}
\label{fig:figure} 
\end{figure}

The production cross sections for the process $p\bar{p}\rightarrow
t\pi_{t}^{0}+X$ at the Tevatron with $\sqrt{s}=1.96TeV$ and the
LHC with $\sqrt{s}=14TeV$ are plotted as functions of the top-pion
mass $m_{\pi_{t}}$ for three values of the free parameter
$\varepsilon$ in Fig.3(a) and Fig.3(b), respectively. From these
figures, we can see that the $t\pi_{t}^{0}$ production cross
section at Tevatron is much smaller than that at LHC in all of the
parameter space of the TC2 models. For $0.02\leq \varepsilon \leq
0.08$ and $200GeV\leq m_{\pi_{t}} \leq500GeV$, the $t\pi_{t}^{0}$
production cross sections at Tevatron and LHC are in the ranges of
$2.3\times10^{-3}fb\sim 4.1fb$ and $17fb\sim 1.82\times10^{3}fb$,
respectively. If we assume the yearly integrated luminosity
$\pounds_{int}=2fb^{-1}$ for the Tevatron with $\sqrt{s}=1.96TeV$,
then the yearly production number of the $t\pi_{t}^{0}$ event is
smaller than 8 in all of the parameter space. Thus, it is very
difficult to detect the possible signals of the neutral top-pion
$\pi_{t}^{0}$ via the process $p\bar{p}\rightarrow t\pi_{t}^{0}+X$
at the Tevatron experiments. However, there will be
$1.7\times10^{3}\sim 1.8\times10^{5}$ $ t\pi_{t}^{0}$ events to be
generated per year at the LHC with $\sqrt{s}=14TeV$ and
$\pounds_{int}=100fb^{-1}$.

\begin{figure}[htb]
\vspace*{-1.5cm}
\begin{center}
\epsfig{file=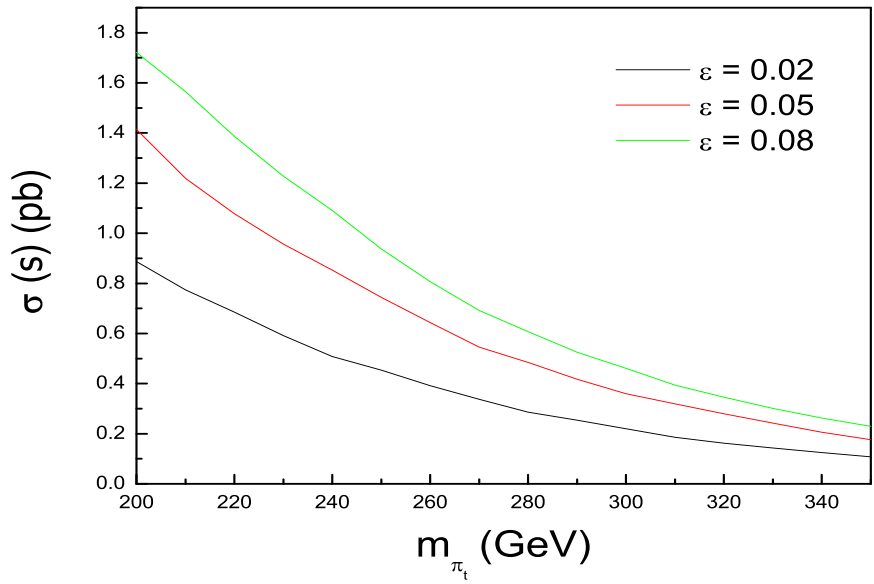,width=370pt,height=270pt} \vspace{-1cm}
\hspace{0.5cm} \caption{The production cross section $\sigma(s)$
of the same sign top pair $tt\bar{c}$ as a function
\hspace*{1.8cm} of $m_{\pi_{t}}$ for three values of the parameter
$\varepsilon$ at the LHC with $\sqrt{s}=14TeV$.} \label{ee}
\end{center}
\end{figure}

The possible decay modes of the neutral top-pion $\pi_{t}^{0}$ are
$t\bar{t},\ \bar{t}c(t\bar{c}), \ b\bar{b},\ gg, \ \gamma\gamma, \
\tau\tau$, and $\mu\mu$. For $m_{t}< m_{\pi_{t}}\leq 2m_{t}$,
$\pi_{t}^{0}$ mainly decays to $\bar{t}c$ or $t\bar{c}$. It has
been shown that the value of the branching ratio
Br($\pi_{t}^{0}\rightarrow \bar{t}c + t\bar{c}$) is larger than
90\% for $m_{\pi_{t}}=250GeV$ and $\varepsilon\geq0.02$[12]. Thus,
for $m_{t}< m_{\pi_{t}}\leq 2m_{t}$, the associated production of
$\pi_{t}^{0}$ with single top quark can easily transfer to the
same sign top pair event $tt\bar{c}$ at LHC. The final state of
the same sign top pair is free from huge QCD background W+jets and
also free from $t\bar{t}$ background[13], which can generate
characteristic signatures at the LHC experiments. Thus,  the same
sign top pair can be used to prob new physics beyond the
SM[14,15,16,17]. So, we further calculate the production cross
section of the same sign top pair final state at LHC. Our
numerical results are shown in Fig.4, in which we have assumed
$m_{t}< m_{\pi_{t}}\leq 2m_{t}$ and taken $\varepsilon=0.02,\
0.05$ and $0.08$. From this figure, we can see that there will be
several and up to ten thousands $tt\bar{c}$ events to be generated
per year at the LHC with $\pounds_{int}=100fb^{-1}$.

 The signal of the same sign top pair event is same
sign dileptons, two b-jets, one charm quark jet plus missing
energy, i.e., $llbbj+\not E$. The main background for this signal
comes from the process $p\overline{p}\rightarrow
W^{\pm}t\bar{t}\rightarrow llbbj_{1}j_{2}+\not E$ with either
$j_{1}$ or $j_{2}$ missing detection. It has been
shown[13,14,15,16] that this background can be significantly
suppressed by applying appropriate cuts and so that the
$tt\bar{c}$ event should be observed at the LHC experiments, as
long as its cross section is larger than several tens $fb$. Thus,
in most of the parameter space of TC2 models, the possible signals
of the neutral top-pion $ \pi_{t}^{0}$ with $m_{t}<
m_{\pi_{t}}\leq 2m_{t}$ can be detected via the process
$p\bar{p}\rightarrow t\pi_{t}^{0}+X\rightarrow tt\bar{c}+X$ at the
LHC experiments.

\begin{figure}[htb]
\vspace{-6.1cm}
\begin{center}
\epsfig{file=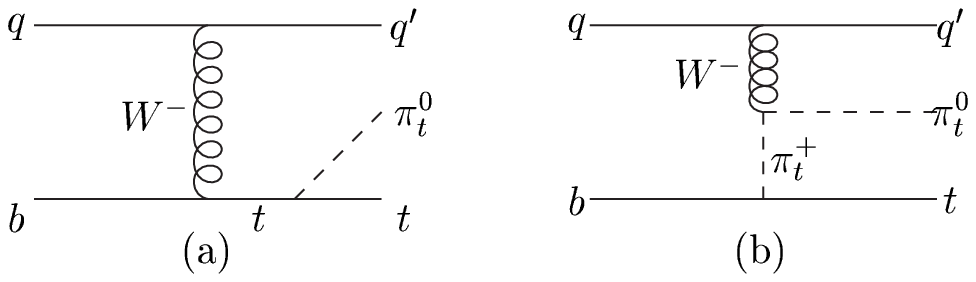,width=550pt,height=800pt} \vspace{-18.5cm}
\hspace{1cm} \vspace{-0.6cm} \caption{Feynman diagrams for the
process $qb\rightarrow q't\pi_{t}^{0}$.} \label{ee}
\end{center}
\end{figure}

In all appearance, the final state which is same as that of
$t\pi_{t}^{0}$ can be reached by single top production followed by
$\pi_{t}^{0}$ bremsstrahlung i.e. the process $qb\rightarrow
q't\pi_{t}^{0}$, as shown in Fig.5. However, due to the unitary
constraint, there exists severe cancellation between Fig.5(a) and
Fig.5(b) and so that its production cross section is highly
suppressed, which is much smaller than that of the subprocess
$cg\rightarrow t\pi_{t}^{0}$[14,16,18]. Thus, this process can not
be taken as an effective process to detect the neutral top-pion
$\pi_{t}^{0}$ at LHC.

For $m_{\pi_{t}}> 2m_{t}$, the neutral top-pion $\pi_{t}^{0}$
mainly decays to $t\bar{t}$ and the associated production of
$\pi_{t}^{0}$ with single top quark can also produce the same sign
top event $tt\bar{t}$ at LHC. However, the signal of this kind of
event is too difficult to extract because of much large
background. Thus, if the neutral top-pion $\pi_{t}^{0}$ is indeed
much heavy, we should consider other processes to detect this type
of new particles in the future high energy experiments.

\noindent{\bf IV. Associated production of the charged top-pions
$\pi_{t}^{\pm}$ with single top quark}
\begin{figure}[htb]
\vspace{-6.9cm}
\begin{center}
\epsfig{file=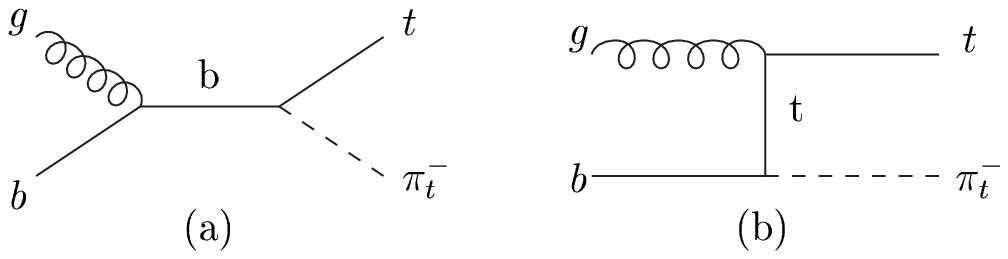,width=550pt,height=800pt} \vspace{-18.5cm}
\hspace{1cm} \vspace{-0.6cm}
 \caption{The Feynman diagrams for the process
$gb\rightarrow t\pi_{t}^{-}$.} \label{ee}
\end{center}
\end{figure}

For TC2 models, the underlying interactions, topcolor
interactions, are assumed to be chiral critically strong at the
scale about 1TeV and coupled preferentially to the third
generation. Thus, top-pions $\pi_{t}^{0,\pm}$ have large Yukawa
couplings to the third family quarks. The charged top-pions
$\pi_{t}^{\pm}$ should be abundantly produced via the subprocess
$gb\rightarrow t\pi_{t}^{\pm}$ at LHC. The relevant Feynman
diagrams are shown in Fig.6.

Our numerical results are shown in Fig.7, in which we plot the
production cross section $\sigma(s)$ for the process
$p\bar{p}\rightarrow t\pi_{t}^{-}+X$ at the Tevatron with
$\sqrt{s}=1.96TeV$[Fig.7(a)] and the LHC with
$\sqrt{s}=14TeV$[Fig.7(b)] as a function of the top-pion mass
$m_{\pi_{t}}$ for three values of the free parameter
$\varepsilon$. One can see from these figures that the cross
section $\sigma(s)$ is not sensitive to the free parameter
$\varepsilon$ and its value at LHC is much large, which is in the
range of $5.24\times 10^{2}fb\sim 2.5\times 10^{4}fb$ for
$0.02\leq \varepsilon \leq 0.08$ and $200GeV\leq m_{\pi_{t}} \leq
500GeV$.

\begin{figure}

\vspace{-0.5cm} \centering {
\label{fig:subfig:a}
\includegraphics[width=3.0in]{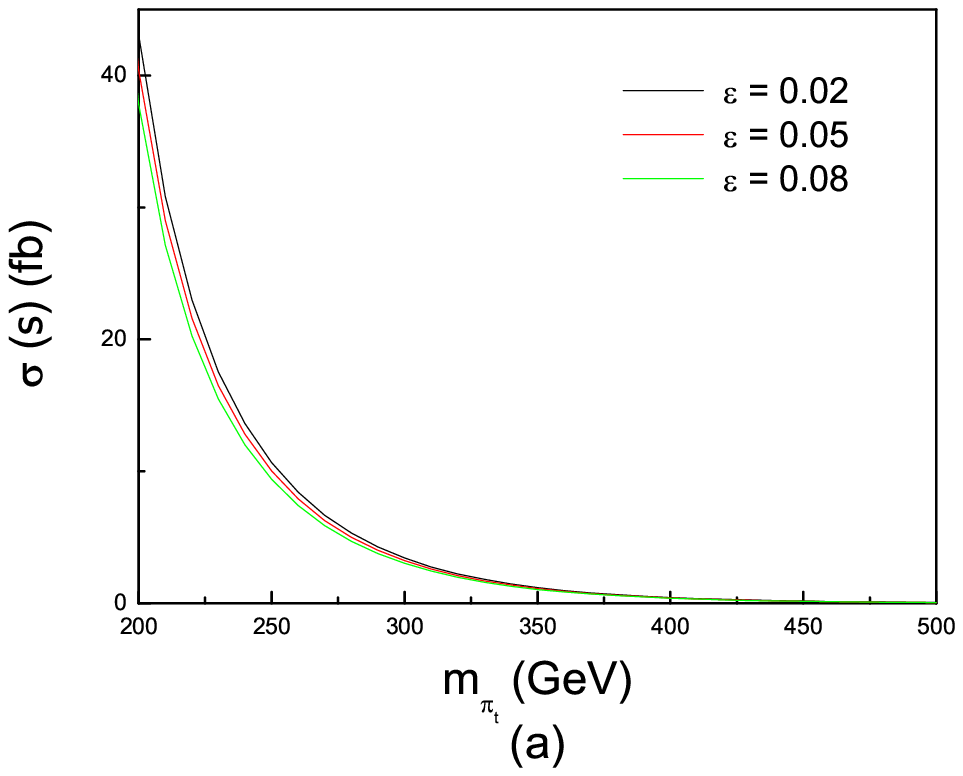}}
\hspace{-0.5cm} \vspace*{0.5cm} {
\label{fig:subfig:b}
\includegraphics[width=3.0in]{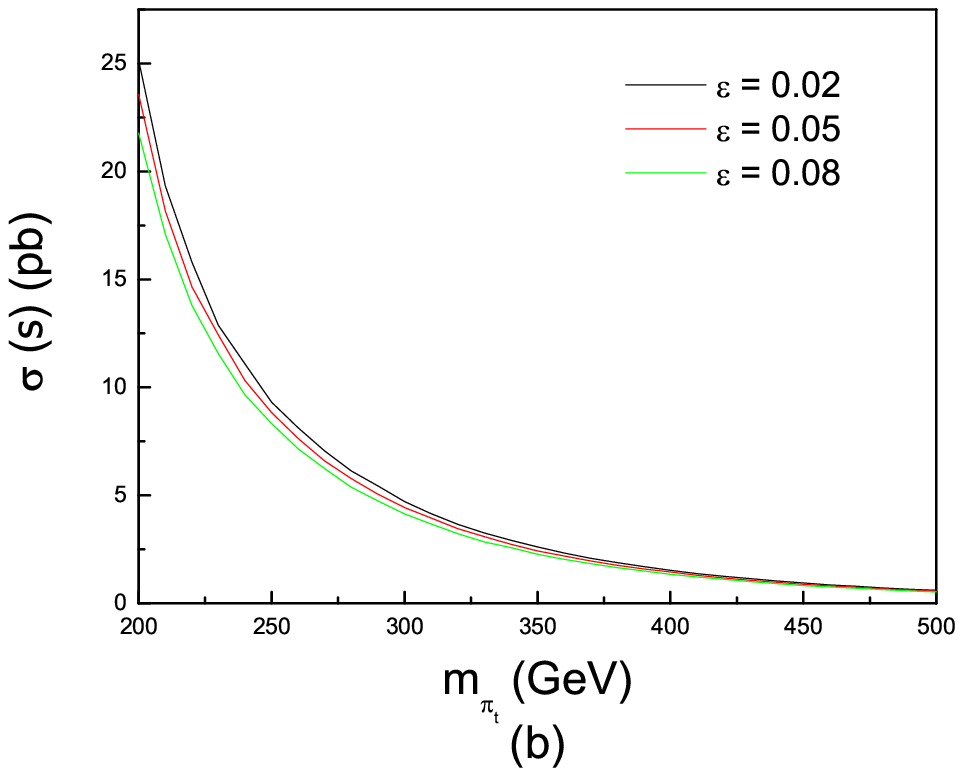}}
\vspace{-0.0cm} \caption{\footnotesize The cross section
$\sigma(s)$ for $t\pi_{t}^{-}$ production as function of
$m_{\pi_{t}}$ for three values of the parameter
\hspace*{1.8cm}$\varepsilon$ at the Tevatron with
$\sqrt{s}=1.96TeV$[Fig.7(a)] and the LHC with
$\sqrt{s}=14TeV$[Fig.7(b)]. }
\label{fig:figure} 

\end{figure}

The LHC has a good potential for discovery of a charged Higgs
boson[19]. Thus, the associated production of the charged Higgs
bosons predicted by the MSSM with single top quark has been
extensively investigated in Refs.[20,21,22]. They have shown that,
considering the complete NLO QCD corrections, the production cross
section for the process $p\bar{p}\rightarrow tH^{\pm}+X$ is
smaller than 1$pb$ in most of the parameter space of MSSM.
Compared with our numerical results, it is smaller than that for
the charged top-pions $\pi_{t}^{\pm}$. This is because the
coupling strength of $H^{\pm}tb$ is smaller than that of
$\pi_{t}^{\pm}tb$.

It has been shown that the heavy Higgs bosons $H^{\pm}$ can be
detected via the decay channels $H^{\pm}\rightarrow
\tau\nu_{\tau}, \ tb$ or $W^{\pm}h^{0}$ at LHC[20,21]. For the
charged top-pions $\pi_{t}^{\pm}$, the dominant decay mode is into
$tb$ channel and its branching ratio is larger than 95$\%$ in most
of the parameter space of TC2 models. It is very difficult to
detect the possible signals of $\pi_{t}^{\pm}$ via the decay
channel $\pi_{t}^{\pm}\rightarrow \tau\nu_{\tau}$. Thus, the
possible signals of $\pi_{t}^{\pm}$ can only be studied via the
process $p p\rightarrow gb\rightarrow t\pi_{t}^{\pm}$ in the $tb$
decay channel. According the analysis results of Ref.[20,21], the
3 b-tags is better for detecting the signals of this process than
the 4 b-tags. For 3 b-tags, the background of the subprocess
$gb\rightarrow t\pi_{t}^{-}\rightarrow t\bar{t}b$ comes from the
NLO QCD processes:
$$gg\rightarrow t\bar{t}b\bar{b} \ \ \ \ \ \ gb\rightarrow t\bar{t}b \ \ \ \ \ \ gg\rightarrow t\bar{t}g$$

After the suitable cuts and the reconstruction of the
$\pi_{t}^{-}$ mass, the value of the signal/background ratio is
large than 5 in most parameter space of TC2 models. Thus, the
charged top-pions $\pi_{t}^{\pm}$ should be observed in the near
future LHC experiments.

\noindent{\bf V. Discussions and conclusions}

The SM predicts the existence of a  neutral Higgs boson, while
many popular models beyond the SM predict the existence of the
neutral or charged scalar particles. These new particles might
produce the observable signatures in the current or future high
energy experiments, which is different from that for the SM Higgs
boson. Any visible signal from the new scalar particles will be
evidence of new physics beyond the SM. Thus, studying the new
scalar particle production at LHC is very interesting.

Topcolor scenario is one of the important candidates for the
mechanism of EWSB. A key feature of this kind of models is that
they predict the existence of the top-pions $\pi_{t}^{0,\pm}$ in
low-energy spectrum. In this paper, we study the associated
production of $\pi_{t}^{0,\pm}$ with single top quark at the
Tevatron with $\sqrt{s}=1.96TeV$ and the LHC with
$\sqrt{s}=14TeV$.

It is well known that the flavor changing neutral current (FCNC)
effects can be used to look for the new physics beyond the SM. The
neutral top-pion $\pi_{t}^{0}$ has large FC coupling to top and
charm quarks at tree-level. Thus, we first calculate the
production cross section of the process $p\bar{p}\rightarrow
t\pi_{t}^{0}+X$ at hadron colliders. We find that the neutral
top-pion $\pi_{t}^{0}$ can be significant generated at the LHC
with $\sqrt{s}=14TeV$. Due to $\pi_{t}^{0}$ mainly decay to
$\bar{t}c$ or $t\bar{c}$, this process can produce a large number
of the same sign top pair $tt\bar{c}$ events. While the production
rates of this kind of events in the SM and the MSSM are far below
the observable level. Thus, we can use the process
$p\bar{p}\rightarrow t\pi_{t}^{0}+X\rightarrow tt\bar{c}+X$ to
look for the neutral top-pion $\pi_{t}^{0}$ at LHC.

For a heavy charged scalar, the dominant production process at LHC
is its associated production with a top quark via gluon bottom
quark fusion. The LHC has good potential for discovering a heavy
charged scalar through this process. In the context of the TC2
models, we calculate the production cross sections of the process
$p\bar{p}\rightarrow t\pi_{t}^{\pm}+X$ at hadron colliders. Our
numerical results show that the production rates are larger than
those for the charged Higgs bosons $H^{\pm}$ from the MSSM. We can
detect the possible signals of the charged top-pions
$\pi_{t}^{\pm}$ at the near future LHC through the process
$p\bar{p}\rightarrow t\pi_{t}^{\pm}+X$ in their $tb$ decay
channel.

All of our numerical results are obtained with the scale chosen
$\mu=2m_{t}$ for the CTEQ6L parton distribution function
integration. Certainly, the numerical results would vary with the
chosen value of the factorization scale varying. For example, if
we chose the scale  $\mu=m_{t}/2$, then $t\pi_{t}^{0}$ production
cross section at LHC is  in the range of $8.4fb\sim
1.8\times10^{3}fb$ for $0.02\leq \varepsilon \leq 0.08$ and
$200GeV\leq m_{\pi_{t}} \leq500GeV$. Comparing with that for the
scale chosen $\mu=2m_{t}$, its value is decreased by $ 10\% \sim
50\% $, which is dependent on the value of the top-pion mass $
m_{\pi_{t}}$. However, even in this case, there will be
$8.4\times10^{2}\sim 1.8\times10^{5} t\pi_{t}^{0}$ events to be
generated per year at the LHC with $\sqrt{s}=14TeV$ and
$\pounds_{int}=100fb^{-1}$, which might be detected in the near
future LHC exprements.

TC2 models also predict the neutral CP-even scalar, called the
top-Higgs boson $h_{t}^{0}$, which is a $t\bar{t}$ bound and
analogous to the $\sigma$ particle in low energy QCD. Similar to
the neutral top-pion $\pi_{t}^{0}$, it also has large coupling to
the top- and charm- quark at tree-level and can give rise to the
anomalous top quark coupling $tcg$. Thus, it can be abundant
produced via the process $p\bar{p}\rightarrow th_{t}^{0}+X$ at
LHC. Our explicit calculation shows that the signal of the
top-Higgs $h_{t}^{0}$ can also be detected through this process in
the near future LHC experiments.

\vspace{0.5cm} \noindent{\bf Acknowledgments}

This work was supported in part by Program for New Century
Excellent Talents in University(NCET-04-0209), the National
Natural Science Foundation of China under the Grants No.10475037.

\end{document}